\documentclass[10pt,twocolumn,twoside,aps,prl,showpacs,superscriptaddress,a4paper]{revtex4-2}
\usepackage{graphicx,epic,eepic,epsfig,amsmath,amsfonts,amsthm,latexsym,amssymb,color,xcolor,enumerate}
\usepackage[colorlinks,linkcolor=blue,anchorcolor=blue,citecolor=blue,]{hyperref}
\newtheorem{theorem}{Theorem}

\newtheorem{corollary}[theorem]{Corollary}

\newtheorem{definition}[theorem]{Definition}

\newtheorem{lemma}[theorem]{Lemma}

\newcommand{\nc}{\newcommand}
\nc{\be}{\begin{equation}}
\nc{\ee}{\end{equation}}
\nc{\ben}{\begin{eqnarray}}
\nc{\een}{\end{eqnarray}}
\nc{\ket}[1]{|#1\rangle}
\nc{\bra}[1]{\langle#1|}
\nc{\braket}[2]{\langle #1 | #2 \rangle}
\nc{\ketbra}[2]{|#1\rangle\!\langle#2|}
\nc{\proj}[1]{| #1\rangle\!\langle #1 |}
\nc{\avg}[1]{\langle#1\rangle}
\nc{\lbar}[1]{\overline{#1}}
\nc{\Rank}{\operatorname{Rank}}
\nc{\smfrac}[2]{\mbox{$\frac{#1}{#2}$}}
\nc{\tr}{\operatorname{Tr}}
\nc{\ox}{\otimes}
\nc{\dg}{\dagger}
\nc{\dn}{\downarrow}
\nc{\supp}{{\operatorname{supp}}}
\nc{\var}{\operatorname{var}}
\nc{\rar}{\rightarrow}
\nc{\lrar}{\longrightarrow}
\nc{\polylog}{\operatorname{polylog}}
\nc{\1}{{\openone}}
\nc{\id}{{\operatorname{id}}}
\nc{\red}[1]{{\textcolor{red}{#1}}}

\nc{\cA}{\mathcal{A}} \nc{\cB}{\mathcal{B}} \nc{\cC}{\mathcal{C}} \nc{\cD}{\mathcal{D}}
\nc{\cE}{\mathcal{E}} \nc{\cF}{\mathcal{F}} \nc{\cG}{\mathcal{G}} \nc{\cH}{\mathcal{H}}
\nc{\cI}{\mathcal{I}} \nc{\cJ}{\mathcal{J}} \nc{\cK}{\mathcal{K}} \nc{\cL}{\mathcal{L}}
\nc{\cM}{\mathcal{M}} \nc{\cN}{\mathcal{N}} \nc{\cO}{\mathcal{O}} \nc{\cP}{\mathcal{P}}
\nc{\cR}{\mathcal{R}} \nc{\cS}{\mathcal{S}} \nc{\cT}{\mathcal{T}} \nc{\cX}{\mathcal{X}}
\nc{\fT}{\mathfrak{T}}

\begin{document}

\title{Reliability Function of Classical-Quantum Channels}

\author{Ke Li}
\email{carl.ke.lee@gmail.com}
\affiliation{Institute for Advanced Study in Mathematics, Harbin Institute of Technology, Harbin 150001, China}

\author{Dong Yang}
\email{yangd6@sustech.edu.cn}
\affiliation{Shenzhen Institute for Quantum Science and Engineering, Southern University of Science and Technology, Shenzhen 518055, China.}
\affiliation{International Quantum Academy, Shenzhen 518048, China}
\affiliation{Guangdong Provincial Key Laboratory of Quantum Science and Engineering, Southern University of Science and Technology, Shenzhen 518055, China}
\date{\today}

\begin{abstract}
We study the reliability function of general classical-quantum channels, which describes the optimal exponent of the decay of decoding error when the communication rate is below the capacity. As the main result, we prove a lower bound, in terms of the quantum R\'enyi information in Petz's form, for the reliability function. This resolves Holevo's conjecture proposed in 2000, a long-standing open problem in quantum information theory. It turns out that the obtained lower bound matches the upper bound derived by Dalai in 2013, when the communication rate is above a critical value. Thus, we have determined the reliability function in this high-rate case. Our approach relies on Renes' breakthrough made in 2022, which relates classical-quantum channel coding to that of privacy amplification, as well as our new characterization of the channel R\'enyi information.
\end{abstract}

\maketitle
Understanding the tradeoff between communication rate and error probability is crucial in the theory of quantum communication~\cite{Wilde-book}. Much work has been done for the problem of transmitting classical information over quantum channels and important results are obtained, including the formula of the capacity~\cite{SW97, Holevo98}, the strong converse property~\cite{Winter99,ON99,KW09} and strong converse exponent~\cite{WWY14, MO17}, higher-order asymptotics~\cite{TT15, CH17, CTT17}, and general characterizations without the assumption of tensor product structure~\cite{HN03,WR12}. Most of these works are concerned with classical-quantum (CQ) channels, a communication model with classical input and quantum output, which captures the essential feature of noncommutativity of quantum mechanics, and at the same time avoids other difficulties such as nonadditivity~\cite{Hastings09, ZZS17} caused by entangled input signals. While the abovementioned topics have been well understood for CQ channels, an important piece---the reliability function, has remained missing as a long-standing open problem.

The investigation of reliability function for CQ channels, and more broadly that in quantum information, dates back to more than two decades ago~\cite{BH, Winter-thesis, Holevo}. For a CQ channel, the reliability function describes the optimal rate of exponential decay of the decoding error with the increase of the number of channel uses, when the communication rate is below the capacity. Thus it provides the precise measure on how rapidly reliable communications can be asymptotically achieved. It is notable that reliability function is a major topic in information theory. For classical channels, it was solved by a series of works of Fano~\cite{Fano}, Gallager~\cite{Gallager}, and Shannon, Gallager and Berlekamp~\cite{SGB}, where the techniques of random coding and sphere packing were introduced to derive lower and upper bounds on the error exponents respectively.

Upper and lower bounds on the reliability function are obtained in the literature for CQ channels. On the one hand, an upper bound in analogy to the classical sphere-packing bound of~\cite{SGB}, was derived by Dalai~\cite{Dalai},  which is believed to be tight. On the other hand, much effort has been devoted to the lower bounds, which concern the achievability part. A tight lower bound was firstly found for the special case where the output states are pure~\cite{BH}. In 2000, Holevo studied general CQ channels and conjectured a random-coding type bound for the achievability of error probability, which implies a lower bound for the reliability function~\cite{Holevo}. Further achievability bounds were provided in Refs.~\cite{Hayashi-1, Hayashi-2, Dalai17, Cheng, BT}. These works apply a direct approach, analysing the decoding error in the single-shot case by use of trace inequalities of operators, which leads to various lower bounds for the error exponent. There is another approach by Renes~\cite{Renes-exponent} that makes use of the relation between channel coding and data compression with quantum side information (DCQSI), as well as the duality relation developed by himself~\cite{Renes-duality} relating DCQSI to quantum privacy amplification. Luckily the error exponent of the latter has already been solved in~\cite{Hayashi-privacy, LYH}. Using this duality technique, a tight lower bound is given for a special class of CQ channels called symmetric channels~\cite{Renes-exponent}.

In this letter, to tackle the problem we follow the duality approach and add a new ingredient based on the type method~\cite{Csiszar}. Combining the two techniques, we eventually resolve Holevo's conjecture~\cite{Holevo} regarding the reliability function in the affirmative. This leads to a tight lower bound to the error exponent, which coincides with the upper bound of Dalai~\cite{Dalai} when the communication rate is above a critical value. Thus, the reliability function of a general CQ channel is determined by combining these two bounds in a proper parameter range.

\emph{Communication with a classical-quantum channel}---A CQ channel $\cN$ is a map from an alphabet $\cX$ to the quantum states space $\cS(B)$ of output system $B$. It sends a classical message $x\in\cX$ to a quantum state $\cN(x)=\rho_x\in\cS(B)$. Given an input distribution $p=(p_x)_x$ on $\cX$, the channel generates a CQ state describing the correlation between the input and the output:
\be\label{eq:cq-state}
\rho_{XB}=\sum_xp_x\ketbra{x}{x}^X\otimes\rho_x^B,
\ee
where $\{\ket{x}^A\}$ is an orthonormal basis. The way to infer the classical message $x$ from the associated output state $\rho_x$ is to perform quantum measurement. Formally, it is represented by a set of positive semidefinite operators $\{\Lambda_x\}_x$ on the Hilbert space of system $B$, such that $\sum_x\Lambda_x=\1_B$ is the identity operator. The probability of obtaining $x'$ from $\rho_x$ is $\tr\rho_x\Lambda_{x'}$. We introduce the optimal decoding error probability for the CQ state of Eq.~\eqref{eq:cq-state}, given by
\be\label{eq:opt-decoding}
P_{\mathrm{err}}(X|B)_{\rho}:=\min_{\{\Lambda_x\}}\left[1-\sum_xp_x\tr\rho_x\Lambda_x\right].
\ee
Equivalently, we also call this the optimal decoding error probability for the input distribution $p$ with respect to the CQ channel. Due to the effect of noise, the states in $\{\rho_x\}$ are usually not distinguishable, thus the error probability does not vanish.

The art to transmit information reliably is to make codes over multiple uses of the channel. For $n$ uses of the channel $\cN$, the input signal is a sequence $x^n:=(x_1,x_2,\ldots,x_n)\in\cX^n$, and  the output state is $\rho_{x^n}:=\rho_{x_1}\otimes\rho_{x_2}\cdots\otimes\rho_{x_n}$. Let $\cM=\{1,...,M\}$ be the set of messages to be transmitted. We select a subset $\cC_n:=\{x^n_1,x^n_2,\ldots,x^n_M\}$ of $\cX^n$ to form a code, where each sequence encodes the corresponding message. Let $\rho^{(n)}_{X^nB^n}$ be the CQ state generated by the channel $\cN^{\ox n}$ corresponding to the \emph{uniform} distribution over the code $\cC_n$. Then the minimum average probability of error for this code is
\be\label{eq:minierr-decoding}
P_e(\cC_n)=P_{\mathrm{err}}(X^n|B^n)_{\rho^{(n)}},
\ee
and the rate of communication is $\frac{1}{n}\log |\cC_n|$, where $|\cC_n|=M$ is the size of the code. Conversely, any CQ state whose classical part is uniformly distributed on a subset of $\cX^n$ defines a code. We are interested in the minimum average probability of error for sending messages at rate $r$ by $n$ uses of the channel $\cN$. Optimised over all the possible codes, it can be written as
\be\label{eq:err-rate}
P_e^{(n)}(\cN,r):=\min\left\{P_e(\cC_n):|\cC_n|\geq 2^{nr}\right\}.
\ee

The channel capacity characterizes the maximal communication rate that the channel $\cN$ can achieve, with asymptotically vanishing error probability. Formally, it is defined as
\be\label{eq:capacity-def}
C(\cN):=\max\left\{r:\lim_{n\to\infty}P_e^{(n)}(\cN,r)=0\right\}.
\ee
The HSW theorem~\cite{SW97, Holevo98} states that the capacity of the CQ channel $\cN$ is given by the Holevo information
\be\label{eq:capacity}
C(\cN)=\max_{p}\left[H\big(\sum_xp_x\rho_x\big)-\sum_xp_xH(\rho_x)\right],
\ee
where $p=(p_x)_x$ is a probability distribution over $\cX$ and $H(\rho)=-\tr(\rho\log\rho)$ is the von Neumann entropy. When $r<C(\cN)$, $P_e^{(n)}(\cN,r)$ is expected to decay to zero exponentially fast. The optimal exponent of this decay, called reliability function, is defined as
\be\label{eq:reliability-def}
E(\cN,r):=\limsup_{n\to\infty}\frac{-1}{n}\log P_e^{(n)}(\cN,r).
\ee

\emph{Petz quantum R\'enyi entropies}---Before stating our main result, we introduce the entropic quantities that we need. Distinct from the classical counterpart, there are more than one version of R\'enyi's information divergence in the quantum setting. The relevant one in the present topic is Petz's version~\cite{Petz86}.

For a quantum state $\rho$ and a positive semidefinite operator $\sigma$ such that the support of the former is contained in that of the latter, the Petz quantum R\'enyi divergence is defined as
\be\label{eq:Renyi-div-def}
D_{\alpha}(\rho\|\sigma):=\frac{1}{\alpha-1}\log\tr\left(\rho^{\alpha}\sigma^{1-\alpha}\right),
~~~0\le\alpha\le 2.
\ee
When $\alpha=1$, we take the limit $\alpha\rar 1$ at the right hand side, and get the quantum relative entropy $D_1(\rho\|\sigma)=D(\rho\|\sigma):=\tr[\rho(\log\rho-\log\sigma)]$. Under this divergence, the quantum R\'enyi mutual information and quantum R\'enyi conditional entropy for a bipartite state $\rho_{AB}$ are given, respectively, as
\ben
I_{\alpha}(A:B)_{\rho} \label{eq:Renyi-mi-def}
&:=&\min_{\sigma_B\in\cS(B)}D_{\alpha}(\rho_{AB}\|\rho_A\otimes\sigma_B), \\
H_{\alpha}(A|B)_{\rho} \label{eq:Renyi-ce-def}
&:=&\max_{\sigma_B\in\cS(B)}\left[-D_\alpha(\rho_{AB}\|\1_A\otimes\sigma_{B})\right].
\een

Now, we extend the notion of R\'enyi mutual information for quantum states to quantum channels. It turns out that this is the right function characterising the reliability function of a general CQ channel.

\begin{definition} \label{def:channel-Renyi-mi}
Let $\cN:\cX\to\cS(B)$ be a CQ channel with $\cN(x)=\rho_x$, and let $p=(p_x)_x$ be a probability distribution on $\cX$. We define
\be\label{eq:p-RenyiI-def}
I_{\alpha}(\cN,p):=I_{\alpha}(X:B)_{\rho}
\ee
with $\rho_{XB}=\sum_xp_x\ketbra{x}{x}^X\otimes\rho_x^B$. Then the R\'enyi information of $\cN$ is defined as
\be\label{eq:Renyi-I-def}
I_{\alpha}(\cN):=\max_{p}I_{\alpha}(\cN,p).
\ee
\end{definition}

\emph{Main results}---The main contribution of this letter is a tight lower bound on the error exponent that has been conjectured for a long time since Holevo's work~\cite{Holevo}.
\begin{theorem}
\label{thm:main}
Let $\cN:\cX\rar\cS(B)$ be a CQ channel and $r\geq 0$. For any $\alpha\in[\frac{1}{2},1]$ and any probability distribution $p$ on the input alphabet $\cX$, we have
\be\label{eq:main-bound}
\limsup_{n\to\infty}\frac{-1}{n}\log P_e^{(n)}(\cN,r)\ge\frac{1-\alpha}{\alpha}[I_{\alpha}(\cN,p)-r].
\ee
\end{theorem}

By combining Theorem~\ref{thm:main} and the upper bound derived in~\cite{Dalai}, we can get the exact exponent of the optimal error probability when the rate of communication is above a critical value.
\begin{theorem}
\label{thm:reliability-function}
Let $\cN:\cX\rar\cS(B)$ be a CQ channel and $r\ge0$. There is a critical value $r_{c}$ such that when $r\ge r_{c}$,
\be\label{eq:reliability}
E(\cN,r)=\max_{\frac{1}{2}\le \alpha\le 1}\frac{1-\alpha}{\alpha}[I_{\alpha}(\cN)-r].
\ee
In general, we have
\ben
E(\cN,r) &&\ge \max_{\frac{1}{2}\le \alpha\le 1}
           \frac{1-\alpha}{\alpha}[I_{\alpha}(\cN)-r],\label{eq:lower-bound}\\
E(\cN,r) &&\le \max_{0 < \alpha\le 1}\frac{1-\alpha}{\alpha}[I_{\alpha}(\cN)-r].
           \label{eq:upper-bound}
\een
\end{theorem}

Eq.~\eqref{eq:lower-bound} results from Theorem~\ref{thm:main} by optimizing the right hand side of Eq.~\eqref{eq:main-bound} over $p$ and $\alpha$. Eq.~\eqref{eq:upper-bound} was proved in~\cite{Dalai}. Eq.~\eqref{eq:reliability} holds because when $r\geq r_c$, the two bounds of Eqs.~\eqref{eq:lower-bound} and \eqref{eq:upper-bound} coincide. From Theorem~\ref{thm:reliability-function}, it can also be seen that the reliability function $E(\cN,r)$ is strictly positive when $r<C(\cN)$, and it is zero when $r\geq C(\cN)$. The arguments for these properties are given in Appendix C.

Our proof of Theorem~\ref{thm:main} relies on two key ideas. The first one is Renes' result on the error exponent of DCQSI, and the fact that in the case of uniformly distributed data, we can construct codes for the corresponding channel from those of DCQSI. The second one is our new finding that the R\'enyi information of a CQ channel associated with an arbitrary input distribution can be asymptotically approximated by that associated with a uniform distribution supported on a subset of the input alphabet. In what follows, we will explain these two ideas first, and then give the proof to Theorem~\ref{thm:main}.

\emph{Data compression with quantum side information}---DCQSI is one of the basic primitives in quantum information processing. Let $\rho_{XB}=\sum_{x}p_x\ketbra{x}{x}^X\otimes\rho_x^B$ be as given in Eq.~\eqref{eq:cq-state}. Here system $X$ at Alice's hands plays the role of the data source, and system $B$ at Bob's hands is the quantum side information. Consider $n$ copies of the state $\rho_{XB}$. The goal of the task is to send a compressed version $\widetilde{X^n}$ of the classical data $X^n$ to Bob, such that on receiving $\widetilde{X^n}$, Bob can reliably recover the original $X^n$.

Renes~\cite{Renes-exponent} derived a tight lower bound on the error exponent, by exploiting the delicate duality relation developed by himself~\cite{Renes-duality} between DCQSI and the task of quantum privacy amplification, whose tight achievability bound was given by Hayashi in~\cite{Hayashi-privacy}. In Renes' compression scheme, the compressed data is generated by dividing $X^n$ into two parts $X^n\cong\widehat{X^n}\widetilde{X^n}$ in an ingenious way, and a copy of $\widetilde{X^n}$ is sent to Bob. Equivalently, Bob only needs to estimate the unsent part $\widehat{X^n}$. The performance of the code is described by the decoding error $P_{\mathrm{err}}\big(\widehat{X^n}|\widetilde{X^n}B^n\big)$, and the fixed compression rate $R_{\mathrm{DC}}:=\frac{1}{n}\log|\widetilde{X^n}|$ that is independent of $n$. Here $|\widetilde{X^n}|$ denotes the dimension of $\widetilde{X^n}$.

\begin{lemma}[Renes~\cite{Renes-exponent}]
\label{lem:Renes-DC}
Consider the task of DCQSI with respect to the CQ state $\rho_{XB}$. For any rate $R_{\mathrm{DC}}>H(X|B)_{\rho_{XB}}$, there is a sequence of compression schemes $X^n\cong\widehat{X^n} \widetilde{X^n}$ with $|\widetilde{X^n}|=2^{nR_{\mathrm{DC}}}$ such that
\ben
&&\lim_{n\to\infty}\frac{-1}{n}\log P_{\mathrm{err}}\left(\widehat{X^n}\big|\widetilde{X^n}B^n\right)_{\rho_{XB}^{\otimes n}} \nonumber\\
&\ge&\max_{\alpha\in[\frac{1}{2},1]}\frac{1-\alpha}{\alpha}
\left[R_{\mathrm{DC}}-H_{\alpha}(X|B)_{\rho_{XB}}\right]. \label{eq:Renes-DC}
\een
\end{lemma}

The relevance of Lemma~\ref{lem:Renes-DC} is that, when $X$ in $\rho_{XB}$ is uniformly distributed, a good compression scheme of DCQSI can be translated into a good code of the channel $\cN$ with $\cN(x)=\rho_x$. This can be easily understood in the one-shot case.  Let $X\cong\widehat{X}\widetilde{X}$ and we rewrite $\rho_{XB}=\rho_{\widehat{X}\widetilde{X}B}=1/(|\widehat{X}||\widetilde{X}|)\sum_{\hat{x},\tilde{x}}
\ketbra{\hat{x}}{\hat{x}}^{\widehat{X}}\otimes\ketbra{\tilde{x}}{\tilde{x}}^{\widetilde{X}}\otimes
\rho_{\hat{x}\tilde{x}}^B$. Then $P_{\mathrm{err}}\big(\widehat{X}|\widetilde{X}B\big)=1/|\widetilde{X}|\sum_{\tilde{x}}P_{\mathrm{err}}
\big(\widehat{X}|\tilde{x}B\big)$, where $P_{\mathrm{err}}\big(\widehat{X}|\tilde{x}B\big)$ are evaluated on states $1/|\widehat{X}|\sum_{\hat{x}}\ketbra{\hat{x}}{\hat{x}}^{\widehat{X}}\otimes\rho_{\hat{x}\tilde{x}}^B$ that represent channel codes of size $|\hat{X}|$ labelled by $\tilde{x}$. Therefore, $P_e^{(1)}(\cN,\log|\widehat{X}|)\le P_{\mathrm{err}}(\widehat{X}|\widetilde{X}B)$. The argument is directly extended to the $n$-copy case, giving
\be\label{eq:DC-Channel}
P_e^{(n)}\left(\cN,\log|\cX|-R_{\mathrm{DC}}\right)
\le P_{\mathrm{err}}\left(\widehat{X^n}|\widetilde{X^n}B^n\right)_{\rho_{XB}^{\otimes n}}.
\ee

Renes used this argument to derive a tight lower bound for the reliability function of certain symmetric CQ channels~\cite{Renes-exponent}. However, for a general CQ channel, the relevant input distribution is usually not uniform, and the above argument does not yield the optimal bound that we want.

\emph{Approximation of R\'enyi information}---We find that the R\'enyi information in Theorem~\ref{thm:main} can be approximated by the R\'enyi information of many copies of the same channel, with an input distribution being uniform over a subset of $\cX^n$. This constitutes the key step that lets us prove Theorem~\ref{thm:main}.

To state the result precisely, we introduce the concept of types. For a sequence $x^n\in\cX^n$, its type $t(x^n)$ is a probability distribution on the alphabet $\cX$, describing the empirical distribution of $x^n$. All the sequences in the set $\cX^n$ of the same type form a type class. So the set $\cX^n$ can be decomposed into disjoint type classes $\cX^n=\bigcup\cT_n$. We use $\frac{\1_{\cT_n}}{|\cT_n|}$ to denote the uniform distribution on the type class $\cT_n$. See Appendix A for the detailed definitions.

\begin{lemma}\label{lem:key}
Let $\cN:\cX\rar\cS(B)$ be a CQ channel. For any $\alpha\in[0,1)\cup(1,2]$ and any probability distribution $p$ on $\cX$, there is a sequence of type classes $\{\cT_n^*\}_n$, such that
\be\label{type}
I_{\alpha}(\cN,p)\le\lim_{n\to\infty}\frac{1}{n}I_{\alpha}\left(\cN^{\otimes n},\frac{\1_{\cT_n^*}}{|\cT_n^*|}\right).
\ee
\end{lemma}

In fact, we can optimize the left hand side of Eq.~\eqref{type} to obtain the following stronger form of an equality. Since we are interested in any probability distribution $p$ on $\cX$, Lemma~\ref{lem:key} will be what we need.

\begin{corollary}\label{cor:key-cor}
Let $\cN:\cX\rar\cS(B)$ be a CQ channel. For any $\alpha\in[0,1)\cup(1,2]$, there is a sequence of type classes $\{\cT_n^*\}_n$, such that
\be\label{eq:key-cor}
I_{\alpha}(\cN)
=\lim_{n\to\infty}\frac{1}{n}I_{\alpha}\left(\cN^{\otimes n},\frac{\1_{\cT_n^*}}{|\cT_n^*|}\right).
\ee
\end{corollary}

The proofs of Lemma~\ref{lem:key} and Corollary~\ref{cor:key-cor} are given in Appendix B.

\emph{Proof of Theorem~\ref{thm:main}}---Let the channel $\cN$ be such that $\cN(x)=\rho_x$. For any integer $m$, consider the state
\be\label{eq:mtproof-1}
\rho_{T_mB^m}^{(m)}:=\sum_{x^m\in\cT_m^*}\frac{1}{|\cT_m^*|}\ketbra{x^m}{x^m}^{T_m}\otimes\rho_{x^m}^{B^m}.
\ee
Here $\cT_m^*$ is the type class given in Lemma~\ref{lem:key}, and $T_m$ is a classical system uniformly distributed on $\cT_m^*$. It can be verified straightforwardly that
\ben
  I_{\alpha}\left(\cN^{\otimes m},\frac{\1_{\cT_m^*}}{|\cT_m^*|}\right)
&=& I_\alpha\left(T_m:B^m\right)_{\rho^{(m)}} \nonumber\\
&=& \log|\cT_m^*|-H_{\alpha}\left(T_m|B^m\right)_{\rho^{(m)}}. \label{eq:mtproof-2}
\een

Consider the task of data compression with respect to the state $\rho_{T_mB^m}^{(m)}$, where the $B^m$ system carries the quantum side information. By Lemma~\ref{lem:Renes-DC}, for any fixed compression rate $R_{\mathrm{DC}}^{(m)}$, there exists a sequence of data compression schemes $T_m^k\cong\widehat{T_m^k}\widetilde{T_m^k}$ with $|\widetilde{T_m^k}|=2^{kR_{\mathrm{DC}}^{(m)}}$ , such that for $\alpha\in[\frac{1}{2},1]$,
\ben
&&\lim_{k\to\infty}\frac{-1}{k}\log P_{\mathrm{err}}
\left(\widehat{T_m^k}\big|\widetilde{T_m^k}B^{mk}\right)_{\left(\rho^{(m)}\right)^{\otimes k}} \nonumber\\
&\ge&\frac{1-\alpha}{\alpha}\left[R_{\mathrm{DC}}^{(m)}-H_{\alpha}\left(T_m|B^m\right)_{\rho^{(m)}}\right].
\label{eq:mtproof-3}
\een

By Eq.~\eqref{eq:DC-Channel}, the above data compression schemes give us a sequence of channel codes for $\cN^{\ox m}$, with communication rate $r^{(m)}=\frac{1}{k}\log|\widehat{T_m^k}|=\log|\cT_m^*|-R_{\mathrm{DC}}^{(m)}$, and at most the same error probability. So,
\ben
&&~~~\limsup_{k\to\infty}\frac{-1}{k}\log P_e^{(k)}\left(\cN^{\otimes m},r^{(m)}\right) \nonumber\\
&&\ge\frac{1-\alpha}{\alpha}\left[R_{\mathrm{DC}}^{(m)}-H_{\alpha}\left(T_m|B^m\right)_{\rho^{(m)}}\right]
   \nonumber\\
&&=\frac{1-\alpha}{\alpha}\left[\log|\cT_m^*|-r^{(m)}
   -H_{\alpha}\left(T_m|B^m\right)_{\rho^{(m)}}\right] \nonumber\\
&&=\frac{1-\alpha}{\alpha}\left[I_{\alpha}\left(\cN^{\otimes m},\frac{\1_{\cT_m^*}}{|\cT_m^*|}\right)
   -r^{(m)}\right]. \label{eq:mtproof-4}
\een

To proceed, we have
\ben
&&~~~\limsup_{n\to\infty}\frac{-1}{n}\log P_e^{(n)}(\cN,r) \nonumber\\
&&\ge\frac{1}{m}\limsup_{k\to\infty}\frac{-1}{k}\log P_e^{(k)}\left(\cN^{\otimes m},mr\right) \nonumber\\
&&\ge\frac{1-\alpha}{\alpha}\left[\frac{1}{m}I_{\alpha}\left(\cN^{\otimes m},
  \frac{\1_{\cT_m^*}}{|\cT_m^*|}\right)-\frac{mr}{m}\right] \nonumber\\
&&\ge\frac{1-\alpha}{\alpha}[I_{\alpha}(\cN,p)-r],  \quad\quad\text{as}\ m\to\infty. \label{eq:mtproof-5}
\een
The first inequality of Eq.~\eqref{eq:mtproof-5} comes from the facts that $P_e^{(mk)}\left(\cN,r\right)=P_e^{(k)}\left(\cN^{\otimes m},mr\right)$ which is a direct consequence of the definition of Eq.~(\ref{eq:err-rate}), and that $\limsup_{n\to\infty}f(n)\ge\limsup_{k\to\infty}f(mk)$ for any function $f$. The last inequality of Eq.~\eqref{eq:mtproof-5} is due to Lemma~\ref{lem:key}.
\qed

\emph{Discussion}---In summary, we have determined the reliability function for a general CQ channel in the case that the communication rate is above a critical value. This is done by proving a tight lower bound, and then combining it with the existing upper bound in the literature. Our result, on the one hand, solves the long-standing open problem on reliability function of general CQ channels, and on the other hand provides an operational interpretation to the R\'enyi information $I_{\alpha}(\cN)$ of Petz's form for a CQ channel $\cN$ in a new parameter range.

A few questions are left for future work. As pointed out already by Renes in~\cite{Renes-exponent}, our proof relies on the solution to DCQSI, which in turn results from its dual problem---privacy amplification. How to derive the lower bound of Theorem~\ref{thm:main} from a direct approach, e.g., by a random-coding argument, remains as an interesting open question.

Another open question is to understand the reliability function when the communication rate is below the critical value. However, this is not solved even for classical channels, which are special CQ channels where the output states are commutative. Indeed, the existence of a critical point in the topic of reliability function is a common phenomenon, and at the unsolved side it usually takes a combinatorial feature and is hard to tackle~\cite{Dalai}.

At last, the R\'enyi entropies are usually used to characterize error exponents, and conversely, the study of error exponents of quantum information tasks lets us identify proper quantum R\'enyi entropies. Therefore, we would like to investigate reliability functions of other quantum information tasks. In particular, it is a natural question whether the result of this letter can be extended to the situation of entanglement-assisted communication over general quantum channels.

\smallskip
\emph{Note added}---After the completion of an earlier version of this work, the authors learned of the work by Renes~\cite{Renes24} which independently obtains the main result of the present letter, by taking a different approach to the problem using Gallager's distribution shaping method.

\bigskip
\emph{Acknowledgements}---The authors would like to thank Hao-Chung Cheng, Aram Harrow and Andreas Winter for valuable discussions on the topic of the present project. They further thank Mark Wilde for comments on an earlier version of the manuscript, and for pointing out improper reference citations. K.L. was supported by the National Natural Science Foundation of China (Grants No. 61871156 and No. 12031004). D.Y. was supported by the National Natural Science Foundation of China (Grant No. 11875244), Guangdong Provincial Key Laboratory (Grant No. 2019B121203002), and the NFR Project (Grant No. ES564777).


\section*{Appendices}

\setcounter{equation}{0}
\renewcommand{\theequation}{A.\arabic{equation}}

\emph{Appendix A: basic concepts and results}---We collect here some fundamental concepts and known results, which are used in the proofs.

The type method~\cite{Csiszar} is a powerful tool in information theory. Let $\cX$ denote a finite alphabet with $|\cX|$ elements. For a sequence $x^n\in\cX^n$, the type $t(x^n):=(t_a(x^n))_a$ is a probability distribution on the alphabet $\cX$, characterizing the empirical distribution of $x^n$ as
\[
t_a(x^n)=\frac{1}{n}\sum_{i=1}^{n}\delta_{x_ia},~~~\forall a\in\cX.
\]
Given a type $t=t(x^n)$ for some sequence $x^n$, the set $\{x^n:x^n\in\cX^n,t(x^n)=t\}$ of sequences of the same type is called a type class. we denote by $\fT_n$ the set of all type classes contained in $\cX^n$, and always use $\cT_n$ to indicate a type class.

With the above notation, $n$ copies of a CQ state $\rho_{XB}=\sum_xp_x\ketbra{x}{x}\otimes\rho_x$  can be expanded as
\be\label{eq:expansion}
\rho_{XB}^{\otimes n}=\sum_{\cT_n\in\fT_n}p^n(\cT_n)\sum_{x^n\in\cT_n}\frac{1}{|\cT_n|}\proj{x^n}\ox\rho_{x^n},
\ee
where
\[
p^n(\cT_n)=\sum_{x^n\in\cT_n}p_{x_1}p_{x_2}\cdots p_{x_n}
\]
and $\rho_{x^n}=\rho_{x_1}\otimes\rho_{x_2}\otimes\cdots\otimes\rho_{x_n}$.

The associated states to the sequences of a fixed type can be transformed to each other by permutation and the average state is a symmetric state. Now we give a very brief overview of symmetric states.

Let $G_n$ be the permutation group over the set $\{1,2,\ldots,n\}$. The natural representation of $G_n$ on $\cH_B^{\otimes n}$ is given by the unitary transformations
\ben
&&V_{\pi}\ket{\phi_1}\otimes\ket{\phi_2}\otimes\cdots\otimes\ket{\phi_n} \nonumber\\
&=&\ket{\phi_{\pi^{-1}(1)}}\otimes\ket{\phi_{\pi^{-1}(2)}}\otimes\cdots\otimes\ket{\phi_{\pi^{-1}(n)}},
\nonumber
\een
for any $\ket{\phi_i}\in\cH_B$ and $\pi\in G_n$. We denote by $\cS_{\mathrm{sym}}(B^n)$ the set of symmetric states of $n$ copies of system $B$, i.e.,
\[
\cS_{\mathrm{sym}}(B^n)
:=\left\{\sigma_{B^n}\!\in\!\cS(B^n):
  \sigma_{B^n}\!=\!V_{\pi}\sigma_{B^n}V_{\pi}^{\dagger},\forall\,\pi\!\in\!G_n\right\}.
\]

For two Hermitian operators $L$ and $K$, we write $L\le K$ if $K-L$ is positive semi-definite. The following lemma gives a useful operator inequality on symmetric states.
\begin{lemma}[\cite{Hayashi-2,CKR}]
\label{lem:uni-sym}
For every finite-dimensional system B and every $n\in \mathbb{N}$, there exists a symmetric state $\sigma_{B^n}^{u}$ such that for any symmetric state $\sigma_{B^n}\in \cS_{\mathrm{sym}}(B^n)$, we have
\[\sigma_{B^n}\le poly(n)\sigma_{B^n}^{u},\]
where $poly(n):=(n+1)^{|B|^2-1}$ is a polynomial of $n$.
\end{lemma}

We remark that such a state is called universal symmetric state and is not unique. The polynomial coefficient can be improved; see~\cite[Lemma 1]{HT} and~\cite[Appendix A]{MO17} for details.

In the following, we list some properties of the Petz quantum R\'enyi entropies.
\begin{lemma}[\cite{MH}]
\label{lem:D-property}
$D_{\alpha}(\rho\|\sigma)$ is convex and nonincreasing with respect to $\sigma$ when $\alpha\in[0,2]$.
\end{lemma}

The monotonicity in Lemma~\ref{lem:D-property} is not shown in~\cite{MH}. Write $D_{\alpha}(\rho\|\sigma)=\frac{1}{\alpha-1}\log\tr\left(\rho^{\frac{\alpha}{2}}\sigma^{1-\alpha}
\rho^{\frac{\alpha}{2}}\right)$. Then for $\alpha\in[0,1)\cup(1,2]$, it follows from the operator monotonicity of the function $x\mapsto x^{1-\alpha}$. For $\alpha=1$, it is well known and is due to the operator monotonicity of the logarithm.

Using quantum Sibson's identity~\cite{Sharma}, the following lemma identifies the unique minimiser in the definition of quantum R\'enyi mutual information of Eq.~(\ref{eq:Renyi-mi-def}).

\begin{lemma}[\cite{GW15}]
\label{lem:Sibson}
For a bipartite state $\rho_{AB}$, we have
\ben
I_{\alpha}(A:B)_{\rho}=D_{\alpha}\left(\rho_{AB}\|\rho_A\otimes\sigma_B^*\right),\nonumber\\
 \text{with}\quad\sigma_B^*:=\frac{\left(\tr_A\rho_A^{1-\alpha}\rho_{AB}^{\alpha}\right)^{\frac{1}{\alpha}}}
 {\tr_B\left[\left(\tr_A\rho_A^{1-\alpha}\rho_{AB}^{\alpha}\right)^{\frac{1}{\alpha}}\right]}. \nonumber
\een
\end{lemma}

Immediately it implies that quantum R\'enyi mutual information is additive for product states. Therefore,
\be\label{eq:add}
I_{\alpha}(A^n:B^n)_{\rho^{\ox n}}=nI_{\alpha}(A:B)_{\rho}.
\ee
By Lemma~\ref{lem:Sibson}, we also have
\be\label{eq:RIforms}
I_{\alpha}(\cN,p)=\frac{\alpha}{\alpha-1}\log\tr\left[
\left(\sum_xp_x\rho_x^{\alpha}\right)^{\frac{1}{\alpha}}\right].
\ee

Equation~\eqref{eq:RIforms} was first shown in~\cite{KW09}, where Lemma~\ref{lem:Sibson} has also been derived in the case that $\rho_{AB}$ is a CQ state. Eq.~\eqref{eq:add} was first proved in~\cite{HT}. Note that the derivation of Lemma~\ref{lem:Sibson} in~\cite{GW15} works for all $\alpha\geq0$, although $\alpha>1$ was imposed there.

The CQ channel's R\'enyi information is additive, and as a function of the order it is continuous and monotonic.
\begin{lemma}[\cite{Holevo}]
\label{lem:Renyi-I-additivity}
For two CQ channels $\cN_1$ and $\cN_2$,
\[
I_{\alpha}(\cN_1\otimes\cN_2)=I_{\alpha}(\cN_1)+I_{\alpha}(\cN_2).
\]
\end{lemma}

\begin{lemma}[\cite{MH, CHT19}]
\label{lem:Renyi-I-monotonicity}
For a CQ channel $\cN$, the function $\alpha\mapsto I_\alpha(\cN)$ is continuous and nondecreasing on $[0,1]$. In particular, $\lim\limits_{\alpha\nearrow 1}I_\alpha(\cN)=C(\cN)$.
\end{lemma}

\setcounter{equation}{0}
\renewcommand{\theequation}{B.\arabic{equation}}

\emph{Appendix B: proof of Lemma~\ref{lem:key} and Corollary~\ref{cor:key-cor}}---At first, we prove in the following Lemma~\ref{lem:unis-approx} a technical result on quantum R\'enyi mutual information for symmetric bipartite states, which will play an important role in the proof of Lemma~\ref{lem:key}.

\begin{lemma}\label{lem:unis-approx}
Let $\rho_{X^nB^n}\in\cS_{\mathrm{sym}}((XB)^n)$, $\sigma_{B^n}^{u}$ and $poly(n)$ be as in Lemma~\ref{lem:uni-sym}. For $\alpha\in[0,2]$ it holds that
\ben
 &&   D_{\alpha}\left(\rho_{X^nB^n}\|\rho_{X^n}\otimes\sigma_{B^n}^{u}\right)-\log poly(n) \nonumber\\
&\le& \min_{\sigma_{B^n}\in\cS(B^n)}D_{\alpha}\left(\rho_{X^nB^n}\|\rho_{X^n}\otimes\sigma_{B^n}\right)  \nonumber\\
&\le& D_{\alpha}\left(\rho_{X^nB^n}\|\rho_{X^n}\otimes\sigma_{B^n}^{u}\right). \nonumber
\een
\end{lemma}

\begin{proof}
The second inequality is obvious. It suffices to prove the first one. We have
\ben
&&\min_{\sigma_{B^n}\in\cS(B^n)}D_{\alpha}(\rho_{X^nB^n}\|\rho_{X^n}\otimes\sigma_{B^n}) \nonumber\\
&\stackrel{(a)}{=}&\min_{\sigma_{B^n}\in\cS_{\mathrm{sym}}(B^n)}D_{\alpha}(\rho_{X^nB^n}\|\rho_{X^n}
 \otimes\sigma_{B^n})  \nonumber\\
&\stackrel{(b)}{\ge}&D_{\alpha}(\rho_{X^nB^n}\|\rho_{X^n}\otimes poly(n)\sigma_{B^n}^u) \nonumber\\
&=&D_{\alpha}(\rho_{X^nB^n}\|\rho_{X^n}\otimes\sigma^u_{B^n})-\log poly(n),  \nonumber
\een
where (b) is by Lemma~\ref{lem:uni-sym} and Lemma~\ref{lem:D-property}. To see (a), due to the invariance of $D_{\alpha}$ under unitary operations, the facts that $\rho_{X^nB^n}\in\cS_{\mathrm{sym}}((XB)^n)$ and $\rho_{X^n}\in\cS_{\mathrm{sym}}(X^n)$, as well as Lemma~\ref{lem:D-property}, we have
\ben
&&D_{\alpha}(\rho_{X^nB^n}\|\rho_{X^n}\otimes\sigma_{B^n}) \nonumber\\
&=&\frac{1}{n!}\sum_{\pi\in\cS_n}
   D_{\alpha}\Big(V_{\pi}^{X^nB^n}\rho_{X^nB^n}V_{\pi}^{\dagger X^nB^n}  \nonumber \\
    &&~~~~~~~~~~~~~~~~\big\|V_{\pi}^{X^nB^n}(\rho_{X^n}\otimes\sigma_{B^n})V_{\pi}^{\dagger X^nB^n}\Big)\nonumber\\
&=&\frac{1}{n!}\sum_{\pi\in\cS_n}D_{\alpha}\left(\rho_{X^nB^n}\big\|\rho_{X^n}\otimes
    V_{\pi}^{B^n}\sigma_{B^n}V_{\pi}^{\dagger B^n}\right) \nonumber\\
&\ge&D_{\alpha}\Big(\rho_{X^nB^n}\big\|\rho_{X^n}\otimes\frac{1}{n!}\sum_{\pi\in\cS_n}
    V_{\pi}^{B^n}\sigma_{B^n}V_{\pi}^{\dagger B^n}\Big). \nonumber
\een
The state $\frac{1}{n!}\sum_{\pi\in\cS_n}V_{\pi}^{B^n}\sigma_{B^n}V_{\pi}^{\dagger B^n}$ is symmetric. This means that the minimization can be restricted to symmetric states.
\end{proof}

It is notable that the above lemma holds for any symmetric state $\rho_{X^nB^n}$, not necessarily CQ.

\smallskip
\begin{proof}[Proof of Lemma~\ref{lem:key}]
Let the channel $\cN$ be such that $\cN(x)=\rho_x$. Write $\rho_{XB}=\sum_xp_x\ketbra{x}{x}^X\otimes\rho_x^B$ and use the expression of Eq.~\eqref{eq:expansion} for the tensor product state $\rho_{XB}^{\ox n}$. We have
\ben
&&I_{\alpha}(\cN,p)  \nonumber\\
&=& \min_{\sigma_B\in\cS(B)}D_\alpha(\rho_{XB}\|\rho_X\ox\sigma_B) \nonumber\\
&\stackrel{(a)}{=}&\frac{1}{n}\min_{\sigma_{B^n}\in\cS(B^n)}
  D_{\alpha}\left(\rho_{XB}^{\otimes n}\|\rho_X^{\otimes n}\otimes\sigma_{B^n}\right) \nonumber\\
&\stackrel{(b)}{\le}&\frac{1}{n}
  D_{\alpha}\left(\rho_{XB}^{\otimes n}\|\rho_X^{\otimes n}\otimes\sigma^u_{B^n}\right) \nonumber\\
&\stackrel{(c)}{=}&\frac{1}{n}\frac{1}{\alpha-1}\log\sum_{\cT_n}p^n(\cT_n)\!\!\sum_{x^n\in\cT_n}\!
  \frac{1}{|\cT_n|}\!\tr\!\left[(\rho_{x^n})^{\alpha}(\sigma^u_{B^n})^{1-\alpha}\right] \nonumber\\
&\stackrel{(d)}{\le}&\frac{1}{n}\frac{1}{\alpha-1}\log\sum_{x^n\in\cT^*_n}\frac{1}{|\cT^*_n|}
  \tr\left[(\rho_{x^n})^{\alpha}(\sigma^u_{B^n})^{1-\alpha}\right]. \label{eq:proof-key-1}
\een
In Eq.~\eqref{eq:proof-key-1}, (a) comes from Lemma~\ref{lem:Sibson}, or more explicitly, Eq.~\eqref{eq:add}, (b) is explicit, (c) is by direct calculation, for (d), noticing that $p^n$ is a probability distribution over type classes, we pick the optimal type class $\cT^*_n$ that minimizes $\sum\limits_{x^n\in\cT_n}\frac{1}{|\cT_n|}
\tr\left[(\rho_{x^n})^{\alpha}(\sigma^u_{B^n})^{1-\alpha}\right]$ when $\alpha\in [0,1)$, and maximizes this quantity when $\alpha\in (1,2]$.

Now, we denote
\[
\rho^{(n)}_{X^nB^n}:=\sum_{x^n\in\cT^*_n}\frac{1}{|\cT^*_n|}\ketbra{x^n}{x^n}^{X^n}\otimes\rho_{x^n}^{B^n}.
\]
Then by the definition of $D_\alpha$, we can easily check that the last line of Eq.~\eqref{eq:proof-key-1} can be written as
\[
\frac{1}{n}D_{\alpha}\left(\rho^{(n)}_{X^nB^n}\big\|\rho^{(n)}_{X^n}\otimes\sigma_{B^n}^{u}\right),
\]
which is further upper bounded as
\ben
&\le&\frac{1}{n}\min_{\sigma_{B^n}\in\cS(B^n)} D_{\alpha}\left(\rho^{(n)}_{X^nB^n}\big\|\rho^{(n)}_{X^n}\otimes\sigma_{B^n}\right)
   +\frac{\log poly(n)}{n} \nonumber\\
&=&\frac{1}{n}I_{\alpha}\left(\cN^{\otimes n},\frac{\1_{\cT_n^*}}{|\cT_n^*|}\right)
   +\frac{\log poly(n)}{n}.\label{eq:proof-key-4}
\een
In Eq.~\eqref{eq:proof-key-4}, the inequality comes from Lemma~\ref{lem:unis-approx}, and the equality is due to Definition~\ref{def:channel-Renyi-mi}.

At last, let $n\to\infty$ and we are done.
\end{proof}

\begin{proof}[Proof of Corollary~\ref{cor:key-cor}]
The $"\le"$ part follows from Lemma~\ref{lem:key}. The $"\ge"$ part is also obvious, as we have
\[
I_{\alpha}\left(\cN^{\otimes n},\frac{\1_{\cT_n^*}}{|\cT_n^*|}\right)
\le I_{\alpha}\left(\cN^{\otimes n}\right)=nI_{\alpha}(\cN),
\]
where the inequality is by Definition~\ref{def:channel-Renyi-mi}, and the equality is by Lemma~\ref{lem:Renyi-I-additivity}.
\end{proof}

\emph{Appendix C: properties of the reliability function}---Lemma~\ref{lem:Renyi-I-monotonicity} and the lower bound of Eq.~\eqref{eq:lower-bound} together imply that $E(\cN,r)$ is strictly positive when $r<C(\cN)$. On the other hand, Lemma~\ref{lem:Renyi-I-monotonicity} and the upper bound of Eq.~\eqref{eq:upper-bound} imply that $E(\cN,r)$ is zero when $r\geq C(\cN)$.

Next, we show Eq.~\eqref{eq:reliability} of Theorem~\ref{thm:reliability-function}. Define
\[
f(\alpha, r)= \frac{1-\alpha}{\alpha}[I_{\alpha}(\cN)-r].
\]
Let $\alpha^*(r)$ be the maximum over all the optimizers of $\max_{0<\alpha\leq 1}f(\alpha,r)$. We claim that: (\romannumeral 1) $\alpha^*(r)=1$ for all $r\geq C(\cN)$, and (\romannumeral 2) $\alpha^*(r)$ is nondecreasing. Claim (\romannumeral 1) is a direct consequence of Lemma~\ref{lem:Renyi-I-monotonicity}. To see (\romannumeral 2), suppose $r_1<r_2\leq C(\cN)$. Then by definition, for any $0<\alpha\leq 1$ we have $f(\alpha^*(r_2),r_2)\geq f(\alpha,r_2)$. Using this inequality and the expression of $f$, we can verify straightforwardly that $f(\alpha^*(r_2),r_1) > f(\alpha,r_1)$ when $\alpha > \alpha^*(r_2)$. This in turn implies that $\alpha^*(r_1)\leq\alpha^*(r_2)$. Now, we set $r_c:=\min\{r:\alpha^*(r)\geq \frac{1}{2}\}$. Then for $r\geq r_c$, we have
\[
\max_{\frac{1}{2}\leq\alpha\leq 1}f(\alpha,r)=\max_{0<\alpha\leq 1}f(\alpha,r),
\]
which leads to Eq.~\eqref{eq:reliability}.


\begin{thebibliography}{99}
\vspace{1mm}

\bibitem{Wilde-book}
M. M. Wilde, \emph{Quantum Information Theory} (Cambridge University Press, 2013).

\bibitem{SW97}
B. Schumacher and M. D. Westmoreland, Sending classical information via noisy quantum channels,
Phys. Rev. A {\bf 56}, 131 (1997).

\bibitem{Holevo98}
A. S. Holevo, The capacity of the quantum channel with general signal states,
IEEE Trans. Inf. Theory {\bf 44}, 269 (1998).

\bibitem{Winter99}
A. Winter, Coding theorem and strong converse for quantum channels,
IEEE Trans. Inf. Theory {\bf 45}, 2481 (1999).

\bibitem{ON99}
T. Ogawa and H. Nagaoka, Strong converse to the quantum channel coding theorem,
IEEE Trans. Inf. Theory {\bf 45}, 2486 (1999).

\bibitem{KW09}
R. K\"{o}nig and S. Wehner, A strong converse for classical channel coding using entangled inputs,
Phys. Rev. Lett. {\bf 103}, 070504 (2009).

\bibitem{WWY14}
M. M. Wilde, A. Winter, and D. Yang, Strong converse for the classical capacity of
entanglement-breaking and Hadamard channels via a sandwiched R\'enyi relative entropy,
Commun. Math. Phys. {\bf 331}, 593 (2014).

\bibitem{MO17}
M. Mosonyi and T. Ogawa, Strong converse exponent for classical-quantum channel coding,
Commun. Math. Phys. {\bf 355}, 373 (2017).

\bibitem{TT15}
M. Tomamichel and V. Y. F. Tan, Second-order asymptotics for the classical capacity
of image-additive quantum channels,
Commun. Math. Phys. {\bf 338}, 103 (2015).

\bibitem{CH17}
H.-C. Cheng and M.-H. Hsieh, Moderate deviation analysis for classical-quantum
channels and quantum hypothesis testing,
IEEE Trans. Inf. Theory {\bf 64}, 1385 (2017).

\bibitem{CTT17}
C. T. Chubb, V. Y. F. Tan, and M. Tomamichel, Moderate deviation analysis for
classical communication over quantum channels,
Commun. Math. Phys. {\bf 355}, 1283 (2017).

\bibitem{HN03}
M. Hayashi, H. Nagaoka, General formulas for capacity of classical-quantum channels,
IEEE Trans. Inf. Theory {\bf 49}, 1753 (2003).

\bibitem{WR12}
L. Wang and R. Renner, One-shot classical-quantum capacity and hypothesis testing,
Phys. Rev. Lett. {\bf 108}, 200501 (2012).

\bibitem{Hastings09}
M. B., Hastings, Superadditivity of communication capacity using entangled inputs,
Nat. Phys. {\bf 5}, 255 (2009).

\bibitem{ZZS17}
E. Y. Zhu, Q. Zhuang, and P. W. Shor, Superadditivity of the classical capacity
with limited entanglement assistance,
Phys. Rev. Lett. {\bf 119}, 040503 (2017).

\bibitem{BH}
M. V. Burnashev and A. S. Holevo, On reliability function of quantum communication channel,
Probl. Inf. Transm. {\bf 34}, 3 (1998).

\bibitem{Winter-thesis}
A. Winter,
Coding theorems of quantum information theory,
PhD Thesis, Universit\"{a}t Bielefeld (1999).

\bibitem{Holevo}
A. S. Holevo, Reliability function of general classical-quantum channel,
IEEE Trans. Inf. Theory {\bf 46}, 2256 (2000).

\bibitem{Fano}
R. M. Fano, \emph{Transmission of Information: A Statistical Theory of Communications}
(M.I.T. Press, 1961).

\bibitem{Gallager}
R. Gallager, A simple derivation of the coding theorem and some applications,
IEEE Trans. Inf. Theory {\bf 11}, 3 (1965).

\bibitem{SGB}
C. Shannon, R. Gallager, and E. Berlekamp, Lower bounds to error probability
for coding on discrete memoryless channels. I,
Inf. Control {\bf 10}, 65 (1967).

\bibitem{Dalai}
M. Dalai, Lower bounds on the probability of error for classical and classical-quantum channels,
IEEE Trans. Inf. Theory {\bf 59}, 8027 (2013).

\bibitem{Hayashi-1}
M. Hayashi, Error exponent in asymmetric quantum hypothesis testing and
its application to classical-quantum channel coding,
Phys. Rev. A {\bf 76}, 062301 (2007).

\bibitem{Hayashi-2}
M. Hayashi, Universal coding for classical-quantum channel,
Commun. Math. Phys. {\bf 289}, 1087 (2009).

\bibitem{Dalai17}
M. Dalai, A note on random coding bounds for classical-quantum channels,
Probl. Inf. Transm. {\bf 53}, 222 (2017).

\bibitem{Cheng}
H.-C. Cheng, A simple and tighter derivation of achievability for classical
communications over quantum channels,
PRX Quantum {\bf 4}, 040330 (2023).

\bibitem{BT}
S. Beigi, M. Tomamichel, Lower bounds on error exponents via a new quantum decoder,
arXiv:2310.09014 (2023).

\bibitem{Renes-exponent}
J. M. Renes, Achievable error exponents of data compression with quantum
side information and communication over symmetric classical-quantum channels,
arXiv:2207.08899 (2022).

\bibitem{Renes-duality}
J. M. Renes, Duality of channels and codes, IEEE Trans. Inf. Theory {\bf 64}, 577 (2018).

\bibitem{Hayashi-privacy}
M. Hayashi, Precise evaluation of leaked information with secure randomness
extraction in the presence of quantum attacker,
Commun. Math. Phys. {\bf 333}, 335 (2015).

\bibitem{LYH}
K. Li, Y. Yao, and M. Hayashi, Tight exponential analysis for smoothing
the max-relative entropy and for quantum privacy amplification,
IEEE Trans. Inf. Theory {\bf 69}, 1680 (2023).

\bibitem{Csiszar}
I. Csiszar, The method of types, IEEE Trans. Inf. Theory {\bf 44}, 2505 (1998).

\bibitem{Petz86}
D. Petz, Quasi-entropies for finite quantum systems,
Rep. Math. Phys. {\bf 23}, 57 (1986).

\bibitem{Renes24}
J. M. Renes, Tight lower bound on the error exponent of classical-quantum channels,
arXiv:2407.11118 (2024).

\bibitem{CKR}
M. Christandl, R. K\"{o}nig, and R. Renner, Postselection technique
for quantum channels with applications to quantum cryptography.
Phys. Rev. Lett. {\bf 102}, 020504 (2009).

\bibitem{HT}
M. Hayashi and M. Tomamichel, Correlation detection and an operational
interpretation of the R\'enyi mutual information,
J. Math. Phys. {\bf 57}, 102201 (2016).

\bibitem{MH}
M. Mosonyi and F. Hiai, On the quantum R\'enyi relative entropies and related capacity formulas,
IEEE Trans. Inf. Theory {\bf 57}, 2474 (2011).

\bibitem{Sharma}
N. Sharma and N. A. Warsi, Fundamental bound on the reliability of quantum information transmission,
Phys. Rev. Lett. {\bf 110}, 080501 (2013).

\bibitem{GW15}
M.~K. Gupta and M.~M. Wilde, Multiplicativity of completely bounded $p$-norms
implies a strong converse for entanglement-assisted capacity,
Commun. Math. Phys. {\bf 334}, 867 (2015).

\bibitem{CHT19}
H.-C. Cheng, M.-H. Hsieh, and M. Tomamichel, Quantum sphere-packing bounds
with polynomial prefactors, IEEE Trans. Inf. Theory {\bf 65}, 2872 (2019).








\end{thebibliography}
\end{document}